# Reciprocity in laser ultrasound revisited: Is wavefield characterisation by scanning laser excitation strictly reciprocal to that by scanning laser detection?

Bernd Köhler[a], Yuui Amano[b], Frank Schubert[a], Kazuyuki Nakahata[b]

[a]Fraunhofer IKTS, Maria-Reiche Str. 2, 01109 Dresden, Germany,
[b]Ehime University, 3 Bunkyo, Matsuyama 790-8577, Japan

Corresponding author: Bernd.koehler@ikts.fraunhofer.de

## Abstract

The common believe about strict measurement reciprocity between scanning laser detection and scanning laser excitation is disproved by a simple experiment. Nevertheless, a deeper study based on the reciprocity relation reveals correct reciprocal measurement set-ups for both the probe-excitation / laser-detection and the laser-excitation / probe-detection case. Similarly, the all-laser measurement, that is thermoelastic laser excitation with laser vibrometer detection, is not in general reciprocal with respect to the exchange of excitation and detection positions. Again, a substitute for the laser doppler vibrometer out-of-plane displacement measurement was found which ensures measurement reciprocity together with laser excitation. The apparent confusion in literature about strict validity/non-validity of measurement reciprocity is mitigated by classifying the measurement situations systematically.

## 1. Introduction

Wavefield visualisation has a long tradition in acoustics and ultrasonic testing. It began as early as the late 18th century with the work done by German physicist August Toepler. He could visualize snapshots of acoustic pulses in transparent fluids by Schlieren methods [1]. Later elastic pulses in transparent solids could be visualized [2, 3]. With the availability of commercial laser doppler vibrometers (LDV), the displacement of the surface of a solid specimen due to an elastic wave could be visualised by scanning techniques [4]. This type of visualization has been very helpful in improving the understanding of the operation of traditional [4] and novel [5-8] types of transducers. Wavefield visualisation also helps to study the propagation of elastic waves in surface graded [9], heterogeneous [4] and fibre composite [10] materials. The dispersion relation [11] and defect interaction [12] of guided waves could also be obtained and even the grain structure of welds could be characterized by detecting waves propagating at grazing incidence [13, 14]. Nowadays, 3D vibrometers also allow the visualisation of the full vector field of the surface displacement [15, 16].

Besides being a tool for basic understanding of elastic wave generation and propagation mechanisms, scanning LDV (SLDV) has been proposed as directly applicable to NDT and SHM [16-19]. However, SLDV is expensive in both the equipment itself as well as the measurement time needed. The latter is due to the massive averaging required caused by the rather poor signal-to-noise ratio of single measurements. Although some speed-up has been achieved by reducing noise through speckle modulation techniques [20, 21] and by increasing the speed of averaging [22], it is unlikely, that SLDV will become established in routine defect detection. Therefore, another recently developed approach deserves closer attention. Instead of exciting ultrasound by a fixed transducer and detecting the wave field on a dense grid of points by SLDV, elastic waves are generated by laser at the same grid points and detected by the same fixed transducer (Scanning Laser Generation Method, SLGM). A reciprocity between SLDV and SLGM is generally assumed [23-25]. Indeed, SLGM provides very good results for many applications [23-29]. It is also applicable to objects with complicated curved surfaces [26]. The biggest advantage of this method is that it is much faster than SLDV, as averaging is usually not required. However, to the best of our knowledge, the equivalence between SLDV and SLGM has neither been theoretically derived nor experimentally demonstrated.

Beside using a piezoelectric sensor as counterpart for laser excitation and detection, respectively, the measurement could also be made fully by lasers by scanning the excitation or the detection laser. It is stated in literature [30] that: "from the linear reciprocity of ultrasonic waves [31, 32] it can be easily shown" that there is a reciprocity for switching the positions of the excitation and detection laser. Unfortunately, an explicit proof was left to the reader, and we could not prove this statement ourself starting from the cited literature ([31, 32]).

The main aim of the present work is therefore to clarify the situation and either find a proof of the claimed reciprocity or alternatively disprove it. This paper is structured as follows. After this introduction we start in Chapter 2 with a taxonomy of measurement reciprocity. In Chapter 3 we describe our experimental set-ups and the numerical methods. This is followed in Chapter 4 by the description of the measurement results for testing the reciprocity between SLGM and SLDV. Analytical derivations are performed in Chapter 5 starting from the foundations of the electromechanical reciprocity theorem and leading to new MRs. In Chapter 6, we use numerical simulation to check the MR for switching the positions between laser excitation and laser detection. Again, the result is supplemented by the search for alternative measurement reciprocities starting from electromechanical theory. In Chapter 7 all results are summarized and discussed.

## 2. Taxonomy of reciprocity

In the literature, the term "reciprocity" is used to describe various things. Reciprocity theorems usually involve differential and integral equations of field variables of two general admissible states of a system. The systems can be very different and involve among others elastostatic, electrostatic, elastodynamic, electrodynamic and coupled elastodynamic-electrodynamic systems [32]. The theorems are helpful in finding analytical solutions to given problems in the corresponding areas, which seems to be the most common application. However, they can also be used to derive strict relations between the results of two different measurements. We call these relations measurement reciprocities (MR). Very often the signal flow is reversed between the two measurements involved in MR. Naturally, the opposite is not true. Not every two measurements, where the signal flow is reversed, do show an MR.

Let us focus on ultrasonic (better denoted as elastodynamic) NDE. As there are different types of measurements, we want to classify them with respect to possible MR. Usually, we have an object to be investigated, a physical mechanism for the generation of an elastodynamic state in the object at a given position and another physical mechanism to record the response at another position. The measurement result is the ratio between the measured output to the exciting input. As MR we denote any fixed relation (mostly the equality) between both ratios.

The most common way to excite elastic waves is by piezoelectric transducers. However, there are many known alternatives, as e.g. electromagnetic acoustic transducers (EMAT), pencil lead break (PLB), laser generated ultrasound or the generation of elastic waves simply by a mechanical (e.g. hammer) impact. A similar variety of possibilities exists for the receiving side. Table 1 gives a collection of transducer possibilities for excitation and detection. Some of them can act both as sender and receiver while others are limited to be either a sender or a receiver.

**Table 1.** Various devices for excitation of ultrasound; physical values that can be used in MR are specified for each transducer type; the symbols are the following. $U$ and $I$: voltage and current at the electrical terminations of the transducer; $f(t)$: normal force to the surface; $f_H(t) = -f_o H(-t)$: a stepwise release of a normal force $-f_o$ at time $t = 0$; $E(t, \tilde{r})$: the laser intensity distribution at the surface; $u, v, a$: displacement, velocity, and acceleration at a surface point, respectively; this table is non-exhaustive, neither for the listing of devices nor for the values which are possibly usable in MR.

| Devices | US probe 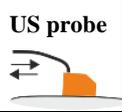 | EMAT 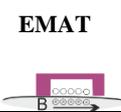 | PLB 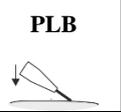 | Hammer / impactor 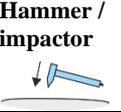 | Laser excitation 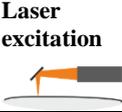 | LDV 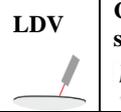 | Capacitive sensor 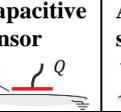 | Acceleration sensor 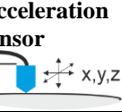 |
|---|---|---|---|---|---|---|---|---|
| **Excitation values** | $U, I$ | $U, I$ | $f_H(t)$ | $f(t)$ | $E(t, \tilde{r})$ | | | |
| **Detection values** | $U, I$ | $U, I$ | | | | $u, v$ | $U, u$ | $U, I, a$ |

Table 1 shows the characteristic physical values for excitation and detection for transducers of different type. The voltage and current of probes are often used as transmitter input and receiver output values. These are the only usable values for piezoelectric and electromagnetic transducers because the values at the transducer object interface are not accessible for measurement. Moreover, theses interface values cannot be directly calculated due to the complex interaction between object and transducer. For other transducers the interaction is non-reactive and the values at the surface can be used directly. We call them "ideal" transducers. In this sense LDV and capacitive transducers are ideal receivers. A lightweight acceleration sensor can be a good approximation to the ideal case. An ideal transmitter for example is the pencil lead break used in acoustic emission.

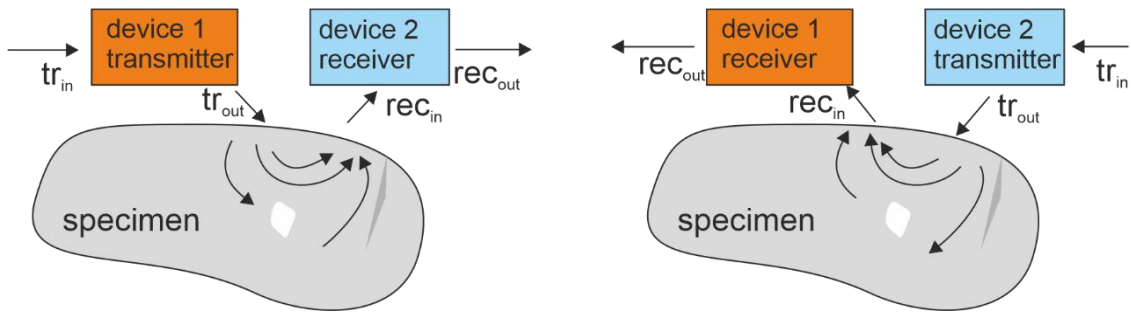

Fig. 1  General scheme of an ultrasonic measurement (left) with the reciprocal measurement situation (right). The signal flow is on the lhs: transmitter input ($tr_{in}$) ➔ transmitter output = specimen input ($tr_{out}$) ➔ specimen output = receiver input ($rec_{in}$) ➔ receiver output ($rec_{out}$). The signal flow on the rhs is reversed. One or both devices can be ideal (non-reactive) transducers.

In deriving MR from the field equations of some reciprocity theorem, usually an integration volume has to be chosen with well-defined boundary conditions. The measurement input and output values are defined at the surface of that volume. For "non-ideal" transducers this volume must contain the transducers as indicated by the dashed line in Fig. 2. Therefore, we have to start with reciprocity theorems for the general coupled elastodynamic – electrodynamic case [40]. When one or both transducers are ideal, they can be excluded from the integration volume (dotted-dashed and dotted lines in Fig. 2). In this case only the values directly on the specimen surface are relevant for the MR, the corresponding transducers can be even different for the transmission and the receiver case. In describing the MR they are not essential and don't have to be specified.

Of course, at this point it is not clear ad-hoc whether there is an MR in the described situations; this must be derived from the measurement theorems in each specific situation.

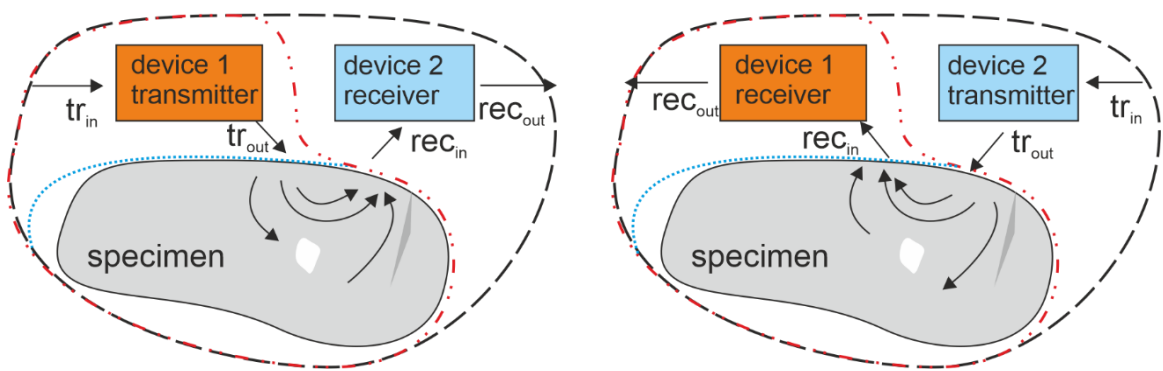

Fig. 2  Copy of Fig. 1 with additional indication of the volumes, where the reciprocity theorems are evaluated. The black dashed line indicates the surface of a volume including the specimen together with both transducers. Applying reciprocity relations to that volume aims to state an MR between the output of the receiver and the input of the sender, usually a voltage or current. The red dot-dashed line encloses the volume when one of the transducers can be considered as ideal (non-reactive) and physical values directly on the surface of the specimen can be used. The blue dotted line is for the case that both transducers are ideal.

| Measurement situation 1 (MS1) | Measurement situation 2 (MS2) | Remarks |
|---|---|---|
| a1) | a2) | a) General measurement reciprocity including probes; proved by Primakoff & Foldy [40] |
| b1) | b2) | b) Ideal transducers for detection and transmission at one of the ports, thus surface values are used there. MR was proven and an experimental validation was performed (see [34]). |
| c1) | c2) | c) The MR with ideal transducers at both ports (for force and velocity) was generally proven [32]; for a recent application see [35]. |
| **Potentially reciprocal measurement situations investigated in this work** | | |
| d1) | d2) | d) One piezoelectric transducer at fixed position combined with a laser excitation/detection. The potential MR between these measurements is evaluated in Chapter 4. |
| e1) | e2) | e) Exchange of the positions in thermoelastic laser excitation and out-of-plane LDV detection. The assumed MR of this constellation is investigated in Chapter 6. |

Fig. 3. Several potential reciprocal measurement situations. MR was proven for the cases a), c) and d). Cases e), f), and g) are the subject of this study. The hollow arrows indicate the direction of the signal flow while the solid arrows describe surface vectors of force and velocity.

Fig. 3 depicts several measurement situations for elastodynamic NDE, some of them with proven MR and some with MR often clamed in literature without proof. The first row includes electromechanical transducers. Primakoff & Foldy [40] could prove for them an MR under the condition that both probes are either pure piezoelectric/electrostatic or pure magnetic/magnetostrictive. Additionally, the excitation must be done by a defined current while the detected value must be the voltage at open contacts.

The MR in the second row could be proven recently [34]. In b1) the surface velocity v(t) is measured and in b2) the excitation by f(t) acts in the same direction. The experimental demonstration was done with a laser vibrometer for velocity detection and the pencil lead break for a definite force step as excitation.

The third example of a proven MR is again long known (see [32] for a general discussion). It is the force excitation at point A with velocity detection at B (c1) which is reciprocal to the same situation with the excitation and detection points interchanged (c2). The component of the velocity must always be taken in the direction in which the force is acting in the complementary measurement.

The last two rows of Fig. 3 represents the main subject of this paper. Each of these rows concerns a pair of measurement situations for which MR is often assumed. The work is about whether this statement is correct. The row d) describes a situation similar to the proven MR given in row b). The difference is in d2) where the excitation is not by a point force as in b2) but by a laser pulse.  Similarly, row e) corresponds to row c) but again the point force of c2) is replaced by the thermoelastic stress generated by a laser source.

## 3.  Methods

### 3.1. Experimental Methods

For experimental verification of the MR in row d) of Fig. 3 we implemented both the piezoelectric excitation with LDV detection and the photoelastic excitation with piezoelectric detection using a similar setup (Fig. 4). The specimen in both cases was the same aluminium plate with a thickness of 30 mm and lateral dimensions of 100 x 100 mm. An ultrasonic probe was coupled to the centre of the top plate surface (Fig. 5). The piezoelectric probe had a diameter of 6.4 mm and a centre frequency of 2.25 MHz. It is designed for excitation and detection of longitudinal bulk waves at zero degree.

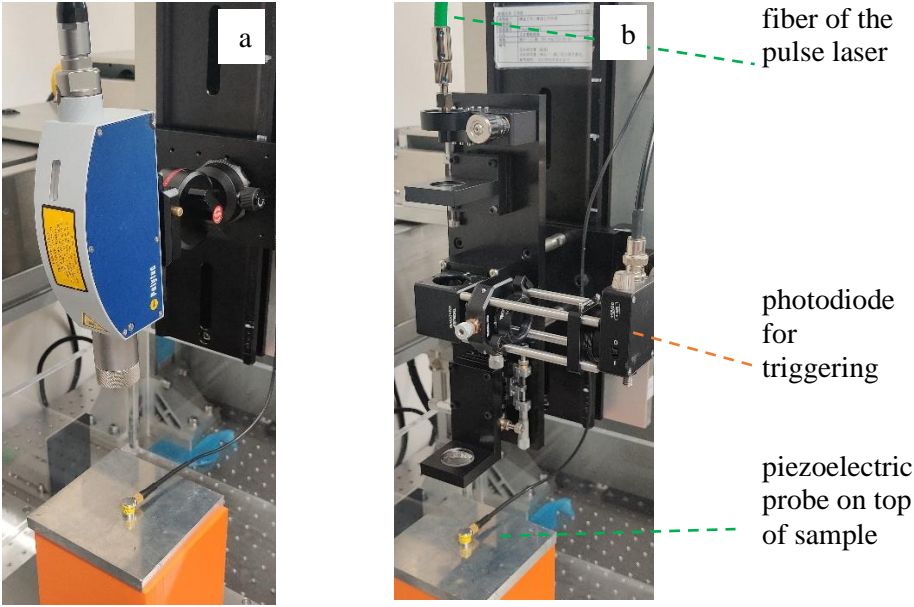

Fig. 4. Experimental setup of the laser heads mounted on the scanner; a: SLDV arrangement with vibrometer head, sample and piezoelectric probe; b: laser excitation arrangement with UV pulse laser barrel with fast photodiode for triggering the data acquisition.

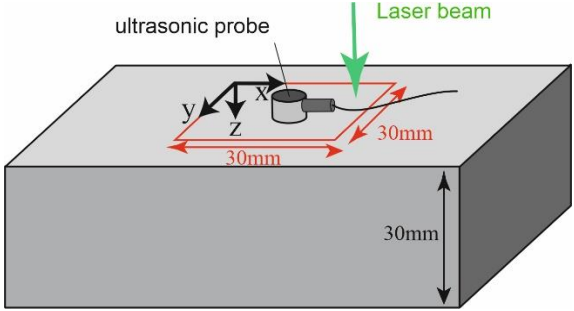

Fig. 5. Aluminium sample with piezoelectric transducer. The scan area of the excitation laser and the LDV probe laser beam is indicated by the red square of 30 x 30 mm².

In the LDV detection experiment the piezoelectric probe acted as transmitter. It was excited by an electrical pulser. The laser vibrometer head was raster scanned over the surface by a mechanical scanner and the out-of-plane displacement was recorded at a grid of points within an area of 30 x 30 mm². The system allows an N-fold averaging of the signals at each given scan point.

For the photoelastic excitation a Litron, Nano L90-100 laser was used. It emits ultraviolet ($\lambda = 532$ nm) light pulses of 4 ns duration with 0.6 mJ energy at a repetition rate of 100 Hz. The laser head (barrel) was mounted on the scanner, focussed to the surface and raster scanned. At each scan point a laser pulse excited elastic waves that were received by the piezoelectric probe. The probe voltage was digitized and stored together with the excitation position.

The trigger of the LDV data acquisition was provided by the electrical pulser of the piezoelectric probe. In the case of photoelastic excitation, a part of the laser beam was decoupled via a beam splitter and directed to a fast photodiode which provided the trigger signal for data acquisition.

### 3.2. Numerical Methods

Various numerical methods can be used to calculate the propagation of elastic waves in solids. Tschöke and Gavenkamp [36] give a good overview of the available methods. One of the rather efficient approaches is the Elastodynamic Finite Integration Technique [37] which is available for 2-D, 3-D and also axisymmetric 2.5D problems. However, in its application to thermoelastic generated elastic waves [38], so far the thermal source was only modelled in a simplified way without taking heat diffusion effects into account. An enhanced, but still two-dimensional EFIT model was recently implemented [39], where the whole coupled thermoelastic problem was included in the FIT modelling. This model was applied here in the numerical investigation of laser generated ultrasound. The laser beam intensity E(x,t) is assumed in the form $E(x,t) = E_0 \, e_1(x) e_2(t)$ with the spatial and time distributions $e_1(x)$ and $e_2(t)$ (Fig. 6). The sample material in the model was aluminium and the following material parameters were used:

longitudinal and transversal sound velocity $\quad c_L = 6.4$ mm/µs, $c_T = 3.15$ µm/µs
specific heat $\quad c = 905$ J/(kg K)
linear expansion coefficient $\quad \alpha = 2.31 \, 10^{-5}$/K
thermal conductivity $\quad k = 237$ W/(mK)

Due to the current restriction of the newly developed thermoelastic EFIT model, all simulations were performed in 2D. That means the sample and the laser source are assumed to extend to infinity in one direction ($x_3$) and all field variables are independent on $x_3$. Irrespective of this limitation it is expected that the 2D plane strain approach will have no influence on the general reciprocity characteristics of the system if compared to a more realistic axisymmetric model.

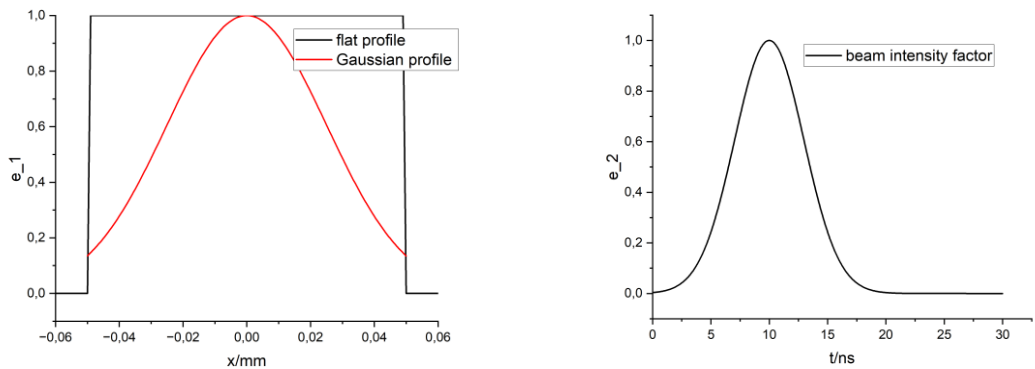

Fig. 6. The laser intensity spatial $e_1(x)$ (left) and temporal $e_2(t)$ (right) intensity distributions. Two versions of spatial distribution were used: one is Gaussian like cutted to 0.1 mm width and the second one is flat, also with 0.1 mm width.

## 4. Visualisation of ultrasonic waves on the sample surface

For both types of measurement, laser excitation and laser detection, raster scans were performed. The pitch of the scan and the scan size were 0.2 mm and 30 x 30 mm², respectively. The number of averaged signals in the LDV detection was N = 400. Wavefront snapshots of both measurement set-ups are shown in Fig. 7. The left column (a) displays the LDV measured surface displacement $u_z$ of the waves generated by the ultrasonic probe. The right column (b) gives corresponding snapshots of the data obtained by the SLGM that is a photoelastic wave generation and detection by the piezoelectric probe.

The circles of the wavefield snapshots are not fully closed, which is a drawback of our measurement setup. The gap in the "3 o´clock" direction is due to the light shielding by the probe cable and the small gaps in the upper and lower part are due to probe fixation stripes. However, these drawbacks do not prevent us to get the essential information from the measurements.

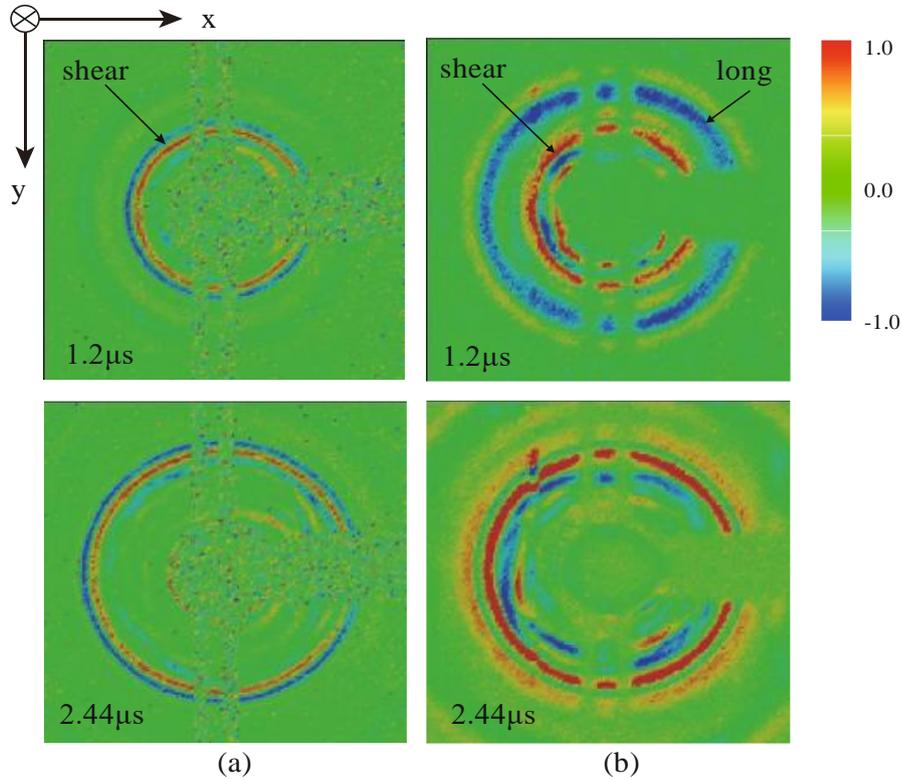

Fig. 7. Snapshots of the recorded data of the SLDV (a, left-hand side) and the SLGM (b, right-hand side). The images on the left-hand side (a) are the recordings of the surface displacement $u_z$ of the probe excited wave, that is, they are direct snapshots of the wave. The right-hand side (b) should be identical to the left-hand side (a) if there is strict reciprocity between laser detection and thermoelastic laser generation (compare Fig. 3 d).

For a given identical scan point, the measurements correspond to measurement situation d1) and d2) of Fig. 3. In our arrangement, the arbitrarily shaped specimen of Fig. 3 specifically has a flat measuring surface. Under the assumption of MR, both signals must be identical at each point and thus the scan snapshots shown in the left and right column of Fig. 7 must also be identical. That is obviously not the case! The thermoelastic excitation case shows an additional wave train at t = 1.2 µs which is not visible in the LDV measurement. Based on the travel path lengths at the snapshot times, the additional wave train could be identified as a skimming longitudinal wave. The wave train which is present in both methods is a shear and/or Rayleigh wave. Most probably it is a superposition of both, since at a short distance from the source the two wave modes are not yet separated. However even this wave train is significantly different in both measurement setups.

With this finding the non-reciprocity of the two measurement situations involving laser detection and laser excitation is clearly shown. The interesting question now is: Can we find a valid reciprocal measurement set-up for each of the two measurement situations? This problem is tackled in the next chapter.

## 5. Exact reciprocity relations

The approach we apply is to specify a reciprocity theorem to the given measurement situation and identify the surface values, which an ideal transducer should measure to fulfil an MR together with the situation of Fig. 3 d1) and another one with Fig. 3 d2). We will start with Fig. 3 d2), the one involving laser excitation.

## 5.1. The reciprocal measurement situation to piezoelectric probe detection of thermoelastically generated waves

### 5.1.1 Description of thermoelastic laser source

To find the reciprocal situation for thermoelastic wave generation, we have to find a mathematical description of the thermoelastic laser source in a form which is appropriate for the application of the reciprocity theorem. The laser pulse is assumed to deposit an energy density $E(\tilde{r})$ into a layer of thickness $d$ at time $t = 0$ (see Fig. 8). The total deposited energy is $\bar{E} = \int E(\tilde{r})dS$, where the integration is extended over the laser illuminated part $S_L$ of the object surface $S$. We assume that the thickness $d$ of the heated layer is small compared to a characteristic lateral scale where the energy density changes significantly. That is, we consider the limit $d \to 0$. We also assume zero heat conductivity.

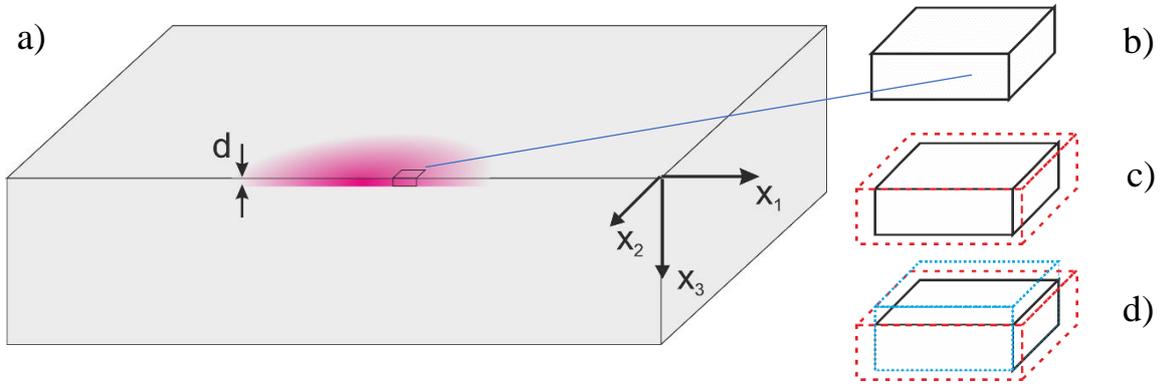

Fig. 8. a) Virtual cut through the sample in the laser spot area. The distribution of the deposited energy is indicated by red colour, b) small piece of thickness d virtually cut out of the surface, c) strain $\epsilon$ in all directions due to free thermal expansion (red dashed shape); there are no stresses in the piece, d) the piece is deformed back to its original lateral dimensions by an additional imposed strain (blue dotted shape), there are stresses in the piece, the tractions necessary for imposing this additional strain are located in the plane $x_3 = d$.

The temperature raise $\Delta T(\tilde{r})$ is assumed constant over the layer depth and amounts to

$$\Delta T(\tilde{r}, t) = \frac{E(\tilde{r})}{d\rho c_V} H(t), \qquad (1)$$

where $\rho$ and $c_V$ are the density and the heat capacity per volume and $H(t)$ is the Heaviside step function. To calculate the stresses in the layer, we adapt and generalize the approach in [33], where the elastic response to the illumination of a disk-shaped surface area with a constant energy density is considered. To be a bit more general, we extend this derivation to an arbitrary lateral distribution of the deposited energy.

To calculate the stress state immediately after the temperature rise, we perform a gedankenexperiment. We consider a cut out element of the layer of thickness d. Further, we assume for a moment it is free and therefore allowed to perform unconstrained ($\sigma_{ij}^{th} = 0 \, \forall i, j$) thermal expansion in all directions, $\epsilon_{ij}^{th} = \alpha_T \Delta T \delta_{ij}$. However, we know the layer is actually clamped laterally as it is inserted in the surface. To bring the layer from the assumed free expansion state back to its actual clamped state we consider the freely expanded layer subjected to an additional imposed strain in **lateral direction** of opposite sign, while keeping the surface normal direction traction free. We call this imposed strain $\bar{\epsilon}$ (Greek indices goes over 1,2):

$$\bar{\epsilon}_{\alpha\beta} = \bar{\epsilon}\delta_{\alpha\beta} \quad \text{with} \quad \bar{\epsilon} = \bar{\epsilon}_{11} = \bar{\epsilon}_{22} = -\alpha_T \Delta T \qquad (2)$$

This ensures that the in-plane components $\epsilon_{\alpha\beta}$ of total strain $\epsilon_{ij} = \epsilon_{ij}^{th} + \bar{\epsilon}_{ij}$ vanish. As the surface is free, we still have $\sigma_{3i} = \bar{\sigma}_{3i} + \sigma_{3i}^{th} = \bar{\sigma}_{3i} = 0$, especially

$$0 = \bar{\sigma}_{33} = (\lambda + 2\mu)\bar{\epsilon}_{33} + 2\lambda\bar{\epsilon}, \qquad (3)$$

and therefore

$$\bar{\epsilon}_{33} = -\frac{2\lambda}{\lambda+2\mu}\bar{\epsilon}. \qquad (4)$$

Note that this strain $\bar{\epsilon}_{33}$ is additional to the strain $\epsilon_{33}^{th}$ generated by the thermal expansion. The total stress in the clamped state is purely due to the imposed strain $\sigma_{ij} = \sigma_{ij}^{th} + \bar{\sigma}_{ij} = \bar{\sigma}_{ij}$:

$$\sigma_{11} = \bar{\sigma}_{11} = (\lambda + 2\mu)\bar{\epsilon}_{11} + \lambda\bar{\epsilon}_{22} + \lambda\bar{\epsilon}_{33} = \frac{2\mu(3\lambda+2\mu)}{\lambda+2\mu}\bar{\epsilon} =: \sigma, \quad \sigma_{22} = \sigma, \quad \sigma_{12} = 0, \qquad (5)$$

where we introduced the variable $\sigma$ (without index) for the in-plane stress components ($\sigma := \sigma_{11} = \sigma_{22}$). This can be written more generally as

$$\sigma_{\alpha\beta} = \sigma\delta_{\alpha\beta}, \quad \sigma_{33} = \sigma_{3\beta} = 0$$

Inserting (1) and (2) into (5) we get

$$\sigma(t,\tilde{r}) = -KH(t)E(\tilde{r})/d \qquad (6)$$

with the dimensionless coupling constant K given by

$$K = \frac{2\mu(3\lambda + 2\mu)}{\lambda + 2\mu}\frac{\alpha_T}{\rho c_V} \qquad (7)$$

In the limit $d \to 0$ we can neglect the inertial forces in the layer:

$$\sigma_{ij,j} = \rho\ddot{u}_i = 0 \qquad (8)$$

from which we get (remember $\sigma_{3\beta} = 0$)

$$\sigma_{\alpha3,3} = -\sigma_{\alpha\beta,\beta} = -\sigma_{,\alpha} \quad \text{and} \quad \sigma_{33,3} = -\sigma_{3\beta,\beta} = 0 \qquad (9)$$

The "in-plane" stress depends on the position in the plane $\tilde{r}$ ($\sigma_{\alpha\beta} = \delta_{\alpha\beta}\sigma(\tilde{r})$) but is constant over the layer thickness (i.e., it does not depend on $x_3$). At the free surface we have no traction ($\sigma_{i3}(x_3 = 0) = 0$). We integrate Equation (9) over the layer thickness and get for the stress in the interface ($x_3 = d$)

$$\sigma_{\alpha3}(d) - 0 = \int_0^d \sigma_{\alpha3,3}\, dx_3 = -\sigma_{,\alpha}d \quad \text{and} \quad \sigma_{33}(d) = 0 \qquad (10)$$

Together with (6) we finally obtain for the traction $\boldsymbol{t} = \boldsymbol{n}\sigma$ exerted from the layer on the underlying material ($x_3 > d$)

$$\boldsymbol{t}(\tilde{r},t) = -K\,H(t)\,\text{grad}^2\big(E(\tilde{r})\big) \qquad (11)$$

Thereby we have defined the traction as usual with respect to the outward normal direction $\boldsymbol{n} = -\boldsymbol{e}_3$ (that is $t_i = -\sigma_{3i}$). The gradient in (11) is taken only in the $x_1 - x_2$ plane which is indicated by the upper index "2" at $\text{grad}^2$. We see that the traction $\boldsymbol{t}$ is purely "in-plane": $\boldsymbol{tn} = 0$. Note that the thickness $d$ of the layer disappeared in the final equation as expected.

To summarize: our model for a thermoelastic pulse laser source of arbitrary spatial energy distribution assumes negligible optical penetration depth and zero heat conductivity. Its wave field can be described as the wave field of a surface traction distribution according to equation (11).

### 5.1.2 Measurement reciprocity

We now aim for the measurement reciprocity. The laser excites an elastic wave at point P which is detected by an electromechanical transducer (probe) converting the laser generated elastic waves to electrical signals (Fig. 9a). A still to be defined electrical value (voltage, current, charge, …) is measured at the transducer output and related to the excitation strength of the laser pulse. The probe acts as receiver so we talk about the receiving transfer function (RTF). Now we try to determine a measurement which is

"reciprocal" in the following sense: The probe is excited electrically and some "value to be defined" is measured at the sample surface position P (Fig. 9b). The measurement should not influence the wave propagation, that is, it should not introduce additional surface tractions. The measured value should be defined such that its ratio to the electrical excitation – the transmission transfer function (TTF) - is identical to the RTF.

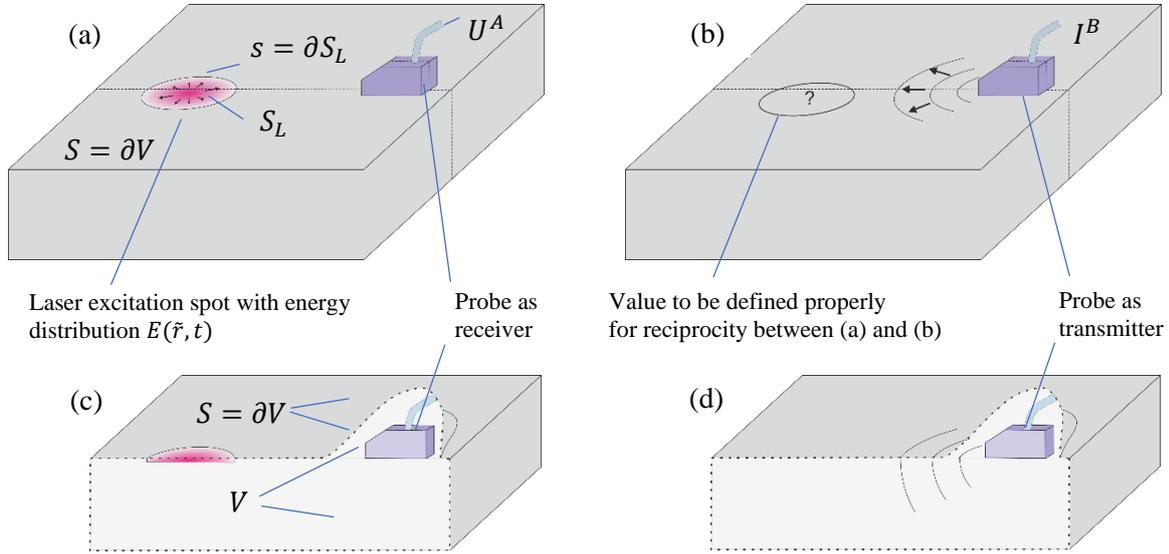

Fig. 9.
(a) The sample for excitation by laser and reception by piezoelectric probe. The laser excitation spot is in the area $S_L$ of the surface S. The red colour indicates the energy distribution $E(\tilde{r})$. The arrows in $S_L$ indicate the surface traction distribution $t(\tilde{r})$ according to (11).
(b) Potentially reciprocal measurement: the probe is active (excited by a current $I^B$). The question mark stands for the quantity that needs to be measured to obtain reciprocity.
(c) The image a) was sectioned at the dotted line to indicate that the volume V includes the sample and the probe. The surface S crosses the coaxial line at some distance from the probe. The voltage U is measured at that crossing.
(d) Corresponding cross section of the image b)
To simplify the drawing, both the laser spot and the probe are drawn at the same flat surface. However, this is not a requirement for the analytical calculations.

We start with the general reciprocity relation as specified in [34] for a sample combined with a piezoelectric probe. A volume V is selected which contains the whole sample and the probe. It is chosen such, that its surface S coincides with the sample surface except in the area around the transducer. Near the transducer the volume surface S is chosen off the sample surface to include the whole transducer and a part of its cable (see Fig. 9c and d). The voltage U and current I are electrical values measured on the coaxial cable where the surface S of the volume V crosses it. The reciprocity relation relates two admissible states of the system denoted by the upper index "A" and "B"

$$\int_{S=\partial V} (t^A \cdot v^B - t^B \cdot v^A) dS + (-)U^B I^A + U^A I^B = 0 \tag{12}$$

where $t^{A,B}$ stands for the surface stresses and $v^{A,B}$ for the surface velocities. Now we specify the state "A" as the laser excitation - probe reception case (Fig. 9a). We measure the voltage at open circuit conditions; there is no current $I^A = 0$. Equation (12) is considered in the frequency domain, so we transform the surface traction (11) from the time into the frequency domain

$$t^A(\tilde{r}, \omega) = -K \hat{H}(\omega) \operatorname{grad}^2(E(\tilde{r})) \tag{13}$$

In the complementary state "B" (Fig. 9b) the probe is active and there are no tractions at the surface S, that is $t^B = 0$ on S. Inserting (13) together with $t^B = 0$ and $I^A = 0$ into (12) the integration over S reduces to an integration over $S_L$ and we get:

$$U^A I^B = K\widehat{H} \int_{S_L} \text{grad}^2(E) \boldsymbol{v}^B dS = K\widehat{H} \int_{S_L} [\text{div}^2(E\boldsymbol{v}^B) - E\text{div}^2\boldsymbol{v}^B] dS \tag{14}$$

We choose the part $S_L$ of the surface S such that all points with $E(\tilde{r}) \neq 0$ are inside $S_L$. Therefore, we have on its border s ($s = \partial S_L$) vanishing energy density ($E = 0$) and by the divergence theorem in 2D, the first term on the right-hand side of (14) vanishes. It remains:

$$U^A I^B = -K\widehat{H} \int_{S_L} [E \, \text{div}^2 \boldsymbol{v}^B] dS. \tag{15}$$

To obtain transfer functions, we first write the laser energy density as $E(\tilde{r}) = \bar{E} E_n(\tilde{r})$ with $\bar{E}$ as the total laser pulse energy in $S_L$ and $\int_{S_L} E_n dS = 1$. Further, we define the value

$$\boldsymbol{u}^B(\boldsymbol{r}, \omega) := \widehat{H}(\omega)\boldsymbol{v}^B(\boldsymbol{r}, \omega) \tag{16}$$

which is essentially the spectrum of the displacement field. Now we can write

$$-\frac{U^A}{K\bar{E}} = \frac{\overline{\text{div} u^B}}{I_B} \tag{17}$$

where the bar on the right-hand side describes the area averaging over $S_L$ with the weighting function $E_n$:

$$\overline{\text{div} u^B} := \int_{S_L} E_n \text{div}^2(\boldsymbol{u}^B) dS = \int_{S_L} E_n \text{div}^2\left(\widehat{H}\boldsymbol{v}^B\right) dS \tag{18}$$

The measurement reciprocity is given by (17) and reads:
    the ratio of the measured value $\overline{\text{div} u^B(\omega)}$ to the excitation current $I^B(\omega)$ (in the probe sending case, Fig. 3d1)
is equal to
    the ratio of the probe voltage response $-U^A(\omega)$ to the laser excitation strength $K\bar{E}$ (Fig. 3d2)

By the derivation above we found a replacement for the v(t) measurement in Fig. 3d1 such that a strict measurement reciprocity (MR) with the laser excitation case d2) is fulfilled. The value to be measured is an (averaged) in-surface strain ($\text{div}^2(\boldsymbol{u})$) given by (18). In most cases this quantity will differ significantly from the out-of-plane velocity, in accordance with the experimental result of non-reciprocity between laser excitation and out-of-plane laser detection. It should be mentioned that for the reciprocity to be valid, the electrical quantities involved are not arbitrary. The piezoelectrically measured quantity must be the open circuit voltage and the electrical excitation quantity must be the current.

The situation simplifies in the limiting case of a **point focused laser**. Setting $E(\tilde{r}) = \bar{E}\delta^2(\tilde{r} - \tilde{r}_P)$, performing the integration over $S_L$ in (15) and transforming the resulting equation back to the time domain we get

$$-U^A(t) * I^B(t) = K\bar{E}H(t) * \text{div}^2 \boldsymbol{v}^B(\tilde{r}, t)|_{\tilde{r}_p} = K\bar{E} \, \text{div}^2 \boldsymbol{u}^B(\tilde{r}, t)|_{\tilde{r}_p} \tag{19}$$

and in the special case of a current pulse $I^B(t) = Q^B \delta(t)$ finally

$$-\frac{U^A(t)}{K\bar{E}} = \frac{\text{div}^2 \boldsymbol{u}^B(\tilde{r}, t)|_{\tilde{r}=\tilde{r}_P}}{Q^B} \tag{20}$$

For the point focused laser excitation the MR now reads:
    the surface strain $\text{div}^2 \boldsymbol{u}^B(\tilde{r}, t)$ measured at $\tilde{r}_p$ and normalized to the current excitation pulse magnitude $Q^B$ of a piezoelectric transducer
is equal to
    the negative of the voltage $U^A(t)$ measured at that same transducer normalized to the laser pulse excitation strength.

Thereby, the laser pulse excitation in case "A" is at the same point $\tilde{r}_p$ at which the strain measurement is performed in case "B" and the transducer voltage measurement is at open contacts ($I^A = 0$).

## 5.2. The reciprocal measurement situation to LDV detection of probe excited waves

In this section we ask for the reciprocal situation to the LDV detection of an elastic wave excited by a piezoelectric transducer. Exactly this situation has been considered in a recent paper of some of the authors where the analysis was based on the work of Primakoff [40], Auld [41] and Achenbach [32]. We therefore refer to [34] and only summarize the results. The relation we are interested in is Eq. 6 of [34]

$$v^A/I^A = -U^B/f^B. \qquad (21)$$

Here $v^A$ is a velocity component of the surface at a given point P due to the excitation of a transducer with a current $I^A$. The component is taken in the direction of a unit vector $e$ determined by the laser beam orientation (Fig. 10a). The voltage $U^B$ is generated at the same transducer for open circuit conditions (no current $I^B = 0$) as a result of a point force excitation $f$ at P with the same orientation $e$ (Fig. 10b). Eq. (21) is taken in the frequency domain but can be transferred to the time domain. In time domain it states the equality of the two values: the velocity response to a current pulse excitation and the open circuit voltage response to a force pulse excitation. Again, the velocity and the force pulse are taken at the same point P with identical direction.

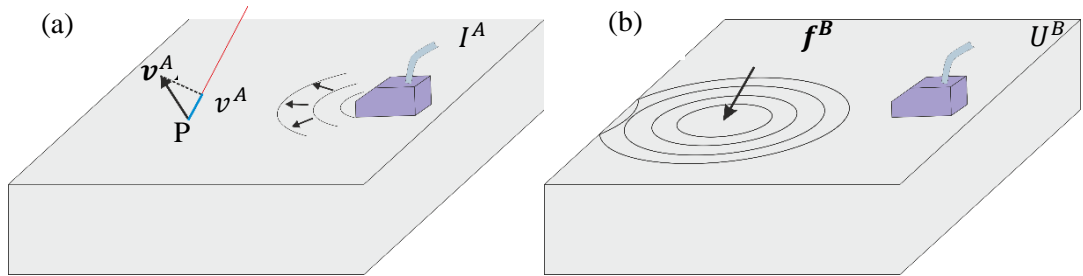

Fig. 10. (a) LDV measurement at P of the waves generated by a piezoelectric probe; the measured value is the projection $v^A$ of the velocity vector $\mathbf{v}^A$ onto the laser beam direction. The electrical excitation of the probe is by the current $I^A$
(b) in the reciprocal measurement the excitation is at the same point P with a point force $\mathbf{f}^B$. The force in (b) has the same orientation as the laser beam in (a). The measured value is the voltage $U^B$ generated by the piezoelectric transducer in open circuit condition ($I^B = 0$).

By LDV usually the out-of-plane component of the velocity is measured. That means that also the point force must act normal to the surface for (21) being valid. A thermoelastic laser source has no normal force component at the surface, so it is not surprising that there is no measurement reciprocity between the two experiments of Chapter 3. However, depending on its operation parameters, such a point force might be generated by a laser ablation source. We know by equation (21) about the exact reciprocity between surface normal (out-of-plane) velocity measurement of probe generated waves and the probe measurements of waves generated by a normal surface force. Therefore, it seems reasonable to use the fulfilment of measurement reciprocity between laser detection and ablation laser generation as an indication, whether the laser ablation source really generates a normal pulse force or whether other surface tractions are also significant. If the normal force dominates, then even a quantitative determination of the force pulse generated by the ablation source seems to be conceivable by using the reciprocal measurement.

## 6. Numerical test of reciprocity between laser excitation and laser detection in full laser ultrasound

Up to now we have investigated the potential reciprocity between the measurement situation in which the laser excitation/detection is combined with a piezoelectric detector/probe. Considering recent research on full laser ultrasonic methods, the question arises whether there is a general measurement reciprocity for exchanging the excitation and detection positions. The fact that we could not prove this type of reciprocity does not mean that it cannot exist. However, one counterexample would disprove it.

Therefore, we check for the simple arrangement of Fig. 11 as a special case of Fig. 3e. The question is whether laser excitation at point P1 and laser vibrometer detection at point P2 gives the same result as laser excitation at P2 and laser detection at P1. We do that by numerical modelling of the laser source and the elastic wave propagation with the extended thermoelastic EFIT model from (see Chapter3.2 ). The modelling was performed in 2D under the assumption that all values depend only on $(x_1, x_2)$ and are constant along the third direction $x_3$.

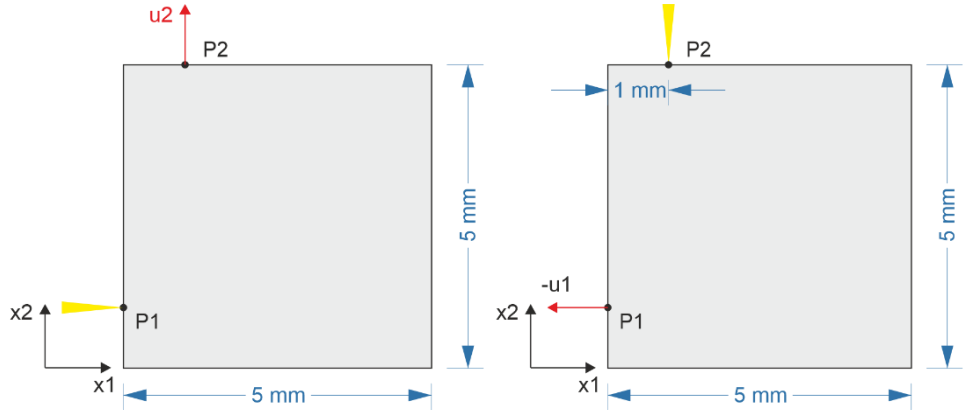

Fig. 11. Two potentially reciprocal situations of laser excitation and out-of-plane (LDV) detection used in the thermoelastic EFIT simulations; a) "Side irradiation": laser excitation at point P1 and displacement detection at point P2; (b) "Top irradiation": laser excitation at P2 and detection at P1.

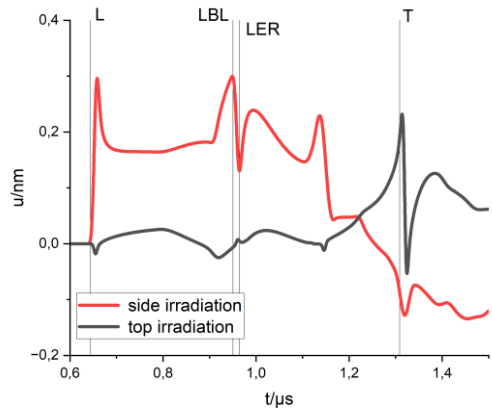

Fig. 12. Out of plane displacement at detection points for laser excitation in both configurations of Fig. 11. The out-of-plane displacement is $u = u_2$ at point P2 (red curve) and $u = -u_1$ at point P1 (black curve). The letters denote the travel times for the following paths: L - direct path of longitudinal wave; LBL (Long – Back wall – Long) - longitudinal wave reflected at the backwall and traveling to the receiving point; T - direct path of the transversal wave; LER (Long – Edge – Rayleigh) – longitudinal wave from P1, mode converted at the upper left edge (E) to an Rayleigh wave (R) and traveling to P2.

The Fig. 12. gives the out-of-plane displacements at the receiving points. There is clearly no MR between these two measurement situations! While the arrival times of various wave modes are still clearly recognisable in both curves their shapes differ considerably. This means that, contrary to what is mentioned in the literature, reciprocity does not generally apply to a position exchange between laser excitation and laser detection.

We learned from the theory in Chapter 5 that reciprocity to laser excitation could be ensured, when the correspondingly detected value is not the out-of-plane displacement, but an appropriate averaged in-plane strain value. There, the counterpart was a reversible piezoelectric transducer. Now we have a similar situation, only the piezoelectric transducer is replaced by also laser excitation and detection. We repeated the analytical considerations for the new situation, and this is presented in the Appendix. The measurement reciprocity is given in (A.10) and the special case of point like excitation is given in (A.12):

$$\overline{\operatorname{div} \mathbf{u}^B}^A = \overline{\operatorname{div} \mathbf{u}^A}^B, \qquad \operatorname{div} \mathbf{u}^B = \operatorname{div} \mathbf{u}^A. \qquad (A.10), (A12)$$

There is reciprocity if the divergence of the surface displacement is taken as the detection value. For a general excitation energy distribution, this divergence must be averaged with the energy distribution of the complementary case, which in indicated by the indexed over-bar in (A.10). For the 2D problems this divergence simplifies to the surface strain that is the derivative of the in-plane displacement.

It is interesting to see whether this analytic result can be confirmed by the numerical simulations. The in-plane-strains at the detection points (P1 and P2) were determined (Fig. 13a). The same laser pulse as for the displacement calculation (Fig. 12) were used. The agreement of the strains is significantly better than the agreement of the displacements. Why is the agreement not perfect as suggested by the reciprocity according to (A.12)? In particular, the signal size at the arrival time of the shear wave differs. The reason is, that the laser pulse is narrow with extension of 0.1 mm, but it is not point like. So, averaged strain according (A.10) must be used instead of the strain at a given point. We repeated the simulation for an intensity distribution that allows an easy calculation of the average. This is a constant laser intensity over a range of 0.1 mm with a sudden drop to zero outside (Fig. 6, black curve). The agreement without averaging is even worth (Fig. 13b). However, with correct averaging (Fig. 13c) the agreement is nearly perfect. The remaining slight deviations are most probably due to numerical errors.

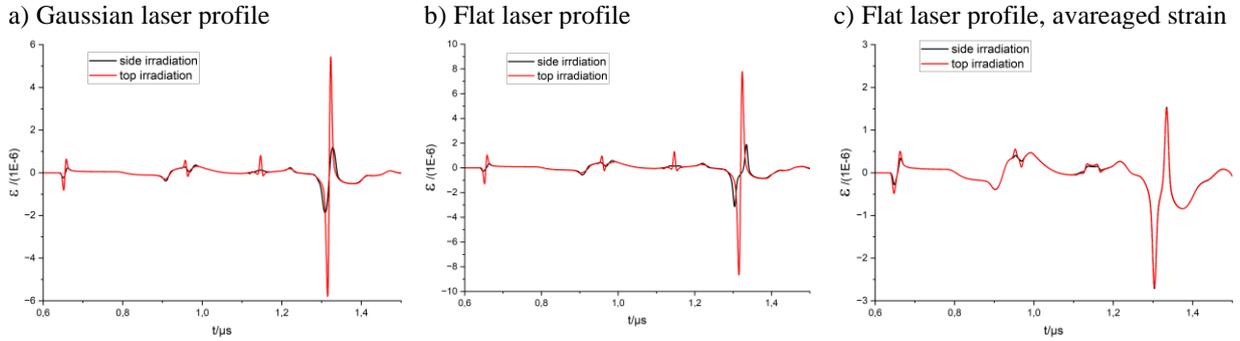

Fig. 13. Simulation results for the in-plane surface strain at the corresponding detection points for top- versus side-irradiation. a) the irradiation has a Gaussian like beam profile; b) the beam profile is constant (flat) over a size of 0.1 mm; c) same as b) however the strain is averaged according to the MR derived in the appendix (equation A.10).

## 7. Summary and Discussions

Visualisation of wavefield propagation has a rather long history with valuable results. Mostly, the waves are generated by a fixed piezoelectric probe and detected by a laser doppler vibrometer scanned over the surface. However, this SLDV has its experimental challenges that limit its widespread application. Most of these challenges can be overcome by replacing the scanning laser detection by scanning the generation laser (SGL) combined with piezoelectric probe detection at a fixed position. To protect the surface from damage, the laser excitation is usually done in the thermoelastic regime. The common believe is, that identical results are obtained between the laser detection and thermoelastic laser excitation at a given point and thus also for the scanning versions SLDV and SLGM. We have shown by a simple experiment, that this is not always the case and that serious deviations from reciprocity can occur.

We studied the reciprocity in laser ultrasound for both situations, asking for the reciprocal situation of thermoelastic laser generated waves detected by a piezoelectric transducer (LGM) and also the reciprocal one to the laser detection of piezoelectric excited waves LDV. To find the reciprocal situation for the LGM is very challenging. For the laser source we assumed instantaneous heating of a thin surface layer and neglected heat conduction. The energy density distribution can be arbitrary. It was shown that this source can be described by pure surface tractions. To the best of our knowledge, this result is new, and we think it could be useful in its own right.

Based on our laser source description we could show that the probe voltage measured for LGM waves is reciprocal to the weighted average surface strain of probe excited waves. Unfortunately, there is currently no obvious practical method to measure surface strains other than strain gauges. They might be appropriate for low frequency experiments with wavelength of (at least) a few millimetres but fail to reach the spatial resolution for typical laser ultrasound experiments. However, technological developments might change this situation in the future. A good candidate for a measuring method is laser speckle photometry [42], although the technique must be expanded for very high image sequences in the MHz range. Further, potential alternatives might be multi path vibrometer based strain measurements [43] and laser optical fibre-bragg-grating measurements [44].

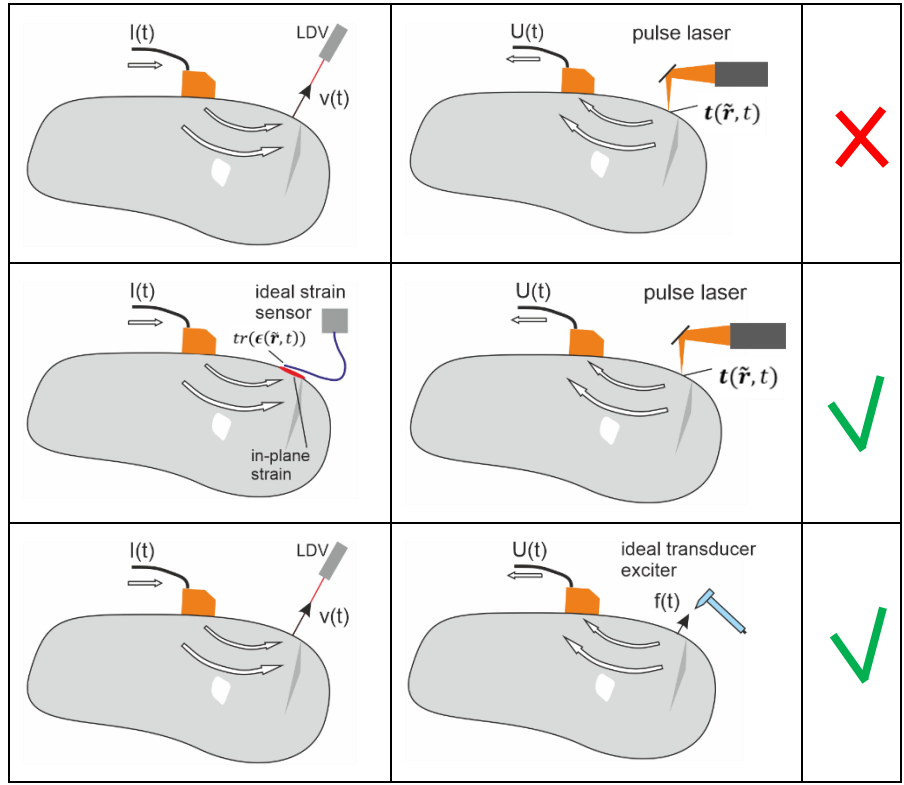

Fig. 14. MR which was demonstrated to be violated (first row) and the value to be measured by some ideal transducer. The reciprocal transducer to a thermoelastic laser source must measure surface strain (second row). To be complementary to surface velocity measurement an ideal transducer must excite a surface force. The last result is already given in Fig. 3b and listed here for completeness.

However, even without an actual practical realisation of a transducer to measure the surface strain, the derived MR has its own value. It tells us what is really mapped in SLGM imaging. This should be very helpful in interpreting these measurements correctly.

The reciprocal situation to the laser detection of the out-of-plane velocity of piezoelectric generated waves was easier to find. It is just the excitation by a normal force. This force can be provided by different ideal transducers. In Fig. 14, an impact hammer is given as example. However also, a laser pulse in the ablation regime produces significant normal force. Therefore, the degree of fulfilment of MR with an ablation laser provides insight whether the source is a normal force only or thermoelastic generated surface tractions contributes too. This way, the degree of MR can tell us something about the source mechanism. The Fig. 14 gives an overview of the measurement pairs with violated MR (red cross) and the newly found pairs with proven MR (green v).

Instead of working with a fixed piezoelectric transducer and scanning the laser beam for either detection or excitation, some authors use a full laser approach. Therefore, the question arises, whether arrangements with laser excitation and laser detection are generally reciprocal to each other with respect to exchange of the positions of both laser beams. That is not the case as we could demonstrate by a simple modelling example. Thus, we disproved the corresponding statement in literature [30]. However, reciprocity is given if the noncontact detection of the out-of-plane displacement by LDV is replaced by some (noncontact) method to measure the in-plane strain. Fig. 15 shows the measurements without MR (red cross) and the potentially pair of measurements with MR (green mark). Again, laser speckle photometry, multi-path vibrometer based strain measurements or laser optical fibre-bragg-grating measurements could probably serve as a method for this in-plain strain measurement in future.

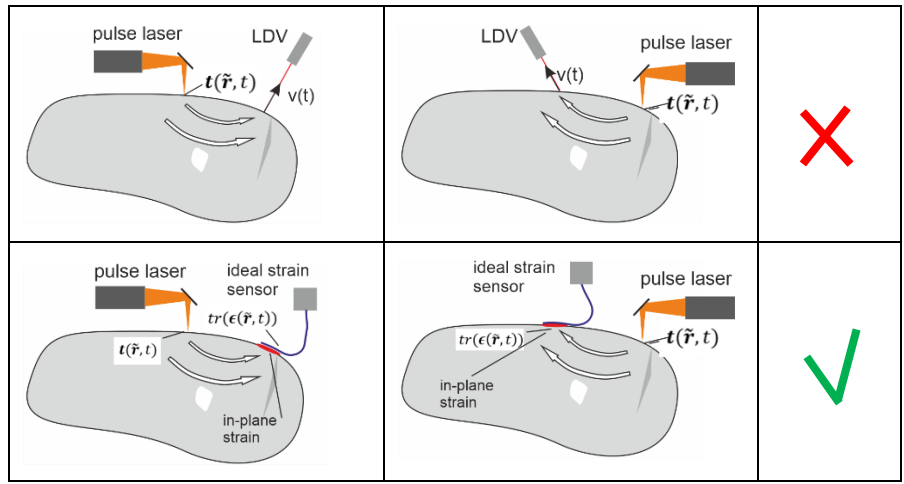

Fig. 15. The MR for switching the positions of thermoelastic laser excitation and LDV detection was demonstrated to be violated (first row). Instead, an ideal transducer should measure the surface strain to provide MR (second row)

Instead of working with a fixed piezoelectric transducer and scanning the laser beam for either detection or excitation, some authors use a full laser approach. Therefore, the question arises, whether arrangements with laser excitation and laser detection are generally reciprocal to each other with respect to exchange of the positions of both laser beams. That is not the case as we could demonstrate by a simple modelling example. Thus, we disproved the corresponding statement in literature [30]. However, reciprocity is given if the noncontact detection of the out-of-plane displacement by LDV is replaced by some (noncontact) method to measure the in-plane strain. Fig. 15 shows the measurements without MR (red cross) and the potentially pair of measurements with MR (green mark). Again, laser speckle photometry, multi-path vibrometer based strain measurements or laser optical fibre-bragg-grating measurements could probably serve as a method for this in-plain strain measurement in future.

Contrary to our results, some publications show an apparent reciprocity for full laser scanning with out-of-plane detection [23]. Obviously, these authors where not interested in the details of the signals but more in characteristic features visible in wavefield snapshots. It might be that the source and receiver characteristics do not play a significant role in those examples and that the wave propagation in the solid itself is reciprocal. However, a detailed discussion of this type of measurement reciprocity is above the scope of the current study.

**Funding:** This work was supported by the JSPS through Invitational Fellowships for Research in Japan (Fellowship ID: S20084).

## References

1. Toepler A, Optische Studien nach der Methode der Schlierenbeobachtung; V. Erscheinungen am elektrischen Funken. Pogg. Ann. Phys. 1868; 132: 194–217 (in German).
2. N. Kudo, H. Ouchi, K. Yamamoto, H. Sekimizu. A simple schlieren system for visualizing a soundfield of pulsed ultrasound, in: *Journal of Physics: Conf. Series*, **1**, 146–9, IOP Publishing, 2004
3. T. Neumann, H. Ermert, "Schlieren visualization of ultrasonic wave fields with high spatial resolution", *Ultrasonics*, **44**, e1561–e1566, 2006, https://doi.org/10.1016/j.ultras.2006.05.209
4. B. Koehler, M. Kehlenbach, R. Bilgam. Optical Measurement and Visualization of Transient Ultrasonic Wave Fields, Proc. of the 27[th] International Acoustical Imaging Symposium, 2003, Saarbücken, Germany, Kluwer Academic Publishers, 2004.

5. Köhler, B., Kim, Y., Chwelatiuk, K., Tschöke, K., Schubert, F., & Schubert, L. (2020). A mode-switchable guided elastic wave transducer. *Journal of Nondestructive Evaluation*, *39*(2), 1-13. https://doi.org/10.1007/s10921-020-00690-5
6. Koehler, B., Frankenstein, B., Schubert, F., & Barth, M. (2009). Novel piezoelectric fiber transducers for mode selective excitation and detection of lamb waves. In *AIP Conference Proceedings* 1096: 982-9. American Institute of Physics. https://doi.org/10.1063/1.3114365
7. Köhler, B., Schubert, F., Barth, M., & Frankenstein, B. (2008). Selective excitation and detection of Lamb waves for SHM applications. In: *Proceedings of the Fourth European Workshop on Structural Health* (pp. 706-14).
8. H. Miao and F. Li. Shear horizontal wave transducers for structural health monitoring and nondestructive testing: A review, Ultrasonics, p. 106355, 2021. https://doi.org/10.1016/j.ultras.2021.106355
9. Ruiz, Alberto, and Peter B. Nagy. Laser-ultrasonic surface wave dispersion measurements on surface-treated metals. *Ultrasonics* 42.1-9 (2004): 665-9. https://doi.org/10.1016/j.ultras.2004.01.045
10. Willberg, C., Koch, S., Mook, G., Pohl, J., & Gabbert, U. (2012). Continuous mode conversion of Lamb waves in CFRP plates. *Smart Materials and Structures*, *21*(7), 075022. doi:10.1088/0964-1726/21/7/075022
11. Köhler, Bernd. Dispersion relations in plate structures studied with a scanning laser vibrometer. *ECNDT, Berlin* (2006).
12. Kehlenbach, M., et al. Numerical and experimental investigation of Lamb wave interaction with discontinuities. *Proceedings of the 4th international workshop on structural health monitoring*. 2003.
13. Köhler, Bernd, et al. Grain structure visualization with surface skimming ultrasonic waves detected by laser vibrometry. *Applied Physics Letters* 101.7 (2012): 074101. http://dx.doi.org/10.1063/1.4745915
14. Kalkowski, M. K., Lowe, M. J., Barth, M., Rjelka, M., & Köhler, B. (2021). How does grazing incidence ultrasonic microscopy work? A study based on grain-scale numerical simulations. *Ultrasonics*, *114*, 106387. https://doi.org/10.1016/j.ultras.2021.106387
15. Barth, Martin, B. Köhler, and L. Schubert. 3D-Visualisation of Lamb waves by laser vibrometry. *Proceedings of the 4th European Workshop on Structural Health Monitoring*. 2008.
16. Staszewski, W. J., Lee, B. C., & Traynor, R. (2007). Fatigue crack detection in metallic structures with Lamb waves and 3D laser vibrometry. *Measurement science and technology*, *18*(3), 727. doi:10.1088/0957-0233/18/3/024
17. Mallet, L., Lee, B. C., Staszewski, J., & Scarpa, F. L. (2003, September). Damage detection in metallic structures using laser acousto-ultrasonics. In *Proceedings of the 4th International Workshop on Structural Health Monitoring, Stanford, CA, USA*.
18. W. J. Staszewski, R. bin Jenal, A. Klepka, M. Szwedo, and T. Uhl. A review of laser Doppler vibrometry for structural health monitoring applications. Key Engineering Materials, vol. 518, pp. 1–15, 2012. doi:10.4028/www.scientific.net/KEM.518.1
19. Sohn, H., Dutta, D., Yang, J. Y., DeSimio, M., Olson, S., & Swenson, E. (2011). Automated detection of delamination and disbond from wavefield images obtained using a scanning laser vibrometer. *Smart Materials and Structures*, *20*(4), 045017. doi:10.1088/0964-1726/20/4/045017
20. Nagy, Peter B., and Gabor Blaho. Random speckle modulation technique for laser interferometry. *Journal of nondestructive evaluation* 11.1 (1992): 41-9.
21. Koehler, B., G. Hentges, and W. Mueller. Improvement of ultrasonic testing of concrete by combining signal conditioning methods, scanning laser vibrometer and space averaging techniques. *NDT & E International* 31.4 (1998): 281-7. https://doi.org/10.1016/S0963-8695(98)00005-X
22. Barth, M. Evaluation device and evaluation method using convolution and deconvolution, (2014), WO2014037414A1.
23. Yashiro, Shigeki, et al. A novel technique for visualizing ultrasonic waves in general solid media by pulsed laser scan. *NDT & E International* 41.2 (2008): 137-44. https://doi.org/10.1016/j.ndteint.2007.08.002
24. Yashiro, S., Toyama, N., Takatsubo, J., & Shiraishi, T. (2010). Laser-Generation Based Imaging of ultrasonic wave propagation on welded steel plates and its application to defect detection. *Materials transactions*, *51*(11), 2069-2075. doi:10.2320/matertrans.M2010204
25. Ostachowicz, W., Radzieński, M., & Kudela, P. (2014). 50th anniversary article: comparison studies of full wavefield signal processing for crack detection. *Strain*, *50*(4), 275-91. https://doi.org/10.1111/str.12098

26. Takatsubo, J., Miyauchi, H., Tsuda, H., Toyama, N., Urabe, K., & Wang, B. (2008). Generation laser scanning method for visualizing ultrasonic waves propagating on 3-D objects. *JOURNAL OF JSNDI*, *57*(4), 162.
27. Sohn, H., Dutta, D., Yang, J. Y., Park, H. J., DeSimio, M., Olson, S., & Swenson, E. (2011). Delamination detection in composites through guided wave field image processing. *Composites science and technology*, *71*(9), 1250-1256. https://doi.org/10.1016/j.compscitech.2011.04.011
28. Mizokami, N., Nakahata, K., Ogi, K., Yamawaki, H., & Shiwa, M. (2017). Numerical simulation of ultrasonic wave propagation in fiber reinforced plastic using image-based modeling. In *AIP Conference Proceedings* (Vol. 1806, No. 1, p. 150005). AIP Publishing LLC. https://doi.org/10.1063/1.4974729
29. Ryuzono, K., Yashiro, S., Nagai, H., & Toyama, N. (2019). Topology optimization-based damage identification using visualized ultrasonic wave propagation. *Materials*, *13*(1), 33. https://doi.org/10.3390/ma13010033
30. An, Y. K., Park, B., & Sohn, H. (2013). Complete noncontact laser ultrasonic imaging for automated crack visualization in a plate. *Smart Materials and Structures*, *22*(2), 025022. DOI 10.1088/0964-1726/22/2/025022
31. Fink, M., and Prada, C. 2001 Acoustic time-reversal mirrors, Inverse Prob. 17 R1–38 DOI 10.1088/0266-5611/17/1/201
32. Achenbach, J.D. Reciprocity in Elastodynamics; Cambridge University Press, Cambridge, 2003
33. Arias, Irene, and Jan D. Achenbach. Thermoelastic generation of ultrasound by line-focused laser irradiation. International journal of solids and structures 40.25 (2003): 6917-35. https://doi.org/10.1016/S0020-7683(03)00345-7
34. Köhler, B., Schubert, L., Barth, M., & Nakahata, K. (2023). Electromechanical Reciprocity Applied to the Sensing Properties of Guided Elastic Wave Transducers. Sensors, 23(1), 150. https://doi.org/10.3390/s23010150
35. Köhler, B., Takahashi, K., & Nakahata, K. (2023). Application of Reciprocity for Facilitation of Wave Field Visualization and Defect Detection. Research in Nondestructive Evaluation, 1-25. https://doi.org/10.1080/09349847.2023.2261878
36. Tschöke, K., and Gravenkamp, H. On the numerical convergence and performance of different spatial discretization techniques for transient elastodynamic wave propagation problems. Wave Motion 82 (2018): 62-85. https://doi.org/10.1016/j.wavemoti.2018.07.002
37. Fellinger, P., Marklein, R., Langenberg, K.J., and Klaholz, S. Numerical modeling of elastic wave propagation and scattering with EFIT - elastodynamic finite integration technique, Wave Motion 21(1), 47-66, (1995). https://doi.org/10.1016/0165-2125(94)00040-C
38. Schubert, F., Peiffer, A., Köhler, B., and Sanderson, T. The elastodynamic finite integration technique for waves in cylindrical geometries. *The Journal of the Acoustical Society of America*, *104*(5), 2604-14 (1998). https://doi.org/10.1121/1.423844
39. Nakahata, K., Miki A., Maruyama. T. Simulation of Photoacoustic Wave Generation and Propagation in Fluid-solid Coupled Media Using Finite Integration Technique, *Proceedings of the 1$^{st}$ Olympiad in Engineering Science – OES2023*
40. Primakoff, H., Foldy, L.L. A General Theory of Passive Linear Electroacoustic Transducers and the Electroacoustic Reciprocity Theorem II. J. Acoust. Soc. Am. 1947, 19, 50-8. https://doi.org/10.1121/1.1916305
41. Auld, B.A. General electromechanical reciprocity relations applied to the calculation of elastic wave scattering coefficients", Wave Motion 1979, 1, 3-10 https://doi.org/10.1016/0165-2125(79)90020-9
42. Cikalova, U., Bendjus, B., Schreiber, J. Laser-Speckle-photometry–A method for non-contact evaluation of material damage, hardness and porosity. 2012 *Materials Testing*, **54**(2), 80-4. https://doi.org/10.3139/120.110299
43. Shambaugh, Kilian, et al. Multi-path Vibrometer-Based Strain Measurement Technique for Very High Cycle Fatigue (VHCF) Testing. *Society for Experimental Mechanics Annual Conference and Exposition*. Cham: Springer Nature Switzerland, 2023.
44. Liu, Guigen, et al. Detection of Fundamental Shear Horizontal Guided Waves Using a Surface-Bonded Chirped Fiber-Bragg-Grating Fabry–Perot Interferometer. *Journal of Lightwave Technology* 36.11 (2018): 2286-94.

# Appendix: A valid measurement reciprocity for full laser ultrasound

The aim is to find a measurement value (detection value) which, together with thermoelastic excitation, fulfils measurement reciprocity in the sense that by exchanging the positions of excitation and detection the same values are measured. The derivation is similar to that of Chapter 5.1. In contrast to the previous considerations, there are no transducers now. The surface of the arbitrarily shaped object is stress-free except in the areas of thermoelastic excitation, i.e. the area of the laser spot. The elastodynamic fields belonging to Fig. 16a and Fig. 16b are denoted with the upper index "A" and "B" respectively.

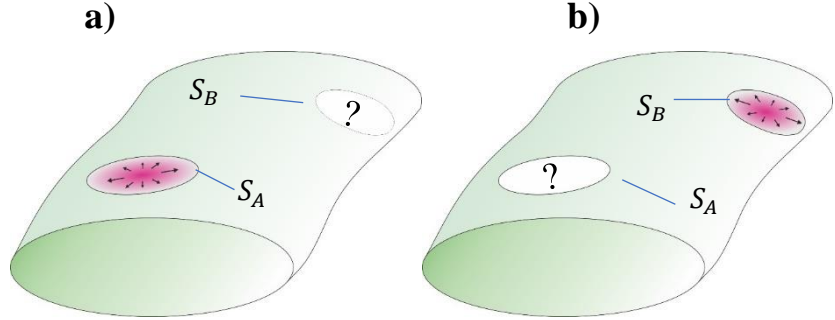

Fig. 16. Arbitrarily shaped surface of an object in two potentially reciprocal measurement situations. The red areas in both situations indicate the surface stress field of thermoelastic excitation. The task is to find those measurement values (indicated by the question mark) which ensure measurement reciprocity.
a): the measurement situation where excitation is at $S_A$ and detection at $S_B$.
b): the reverse situation with excitation at $S_B$.

We again assume that the energy density $E(\tilde{r}) = \bar{E} E_n(\tilde{r})$ is deposited instantly in the excitation area where $\bar{E}$ is the total energy and $E_n(\tilde{r})$ is the normalised energy density. To be general, we allow different total energies and distributions

$$E^{A/B}(\tilde{r}) = \bar{E}^{A/B} E_n^{A/B}(\tilde{r}) \tag{A.1}$$

for both considered states.

Instead of (12) we have now the simpler relation

$$\int_{S=\partial V} (\boldsymbol{t}^A \cdot \boldsymbol{v}^B - \boldsymbol{t}^B \cdot \boldsymbol{v}^A) dS = 0 \tag{A.2}$$

which can be split in two integrals and restricted to the areas of nonvanishing tractions giving

$$\int_{S_A} \boldsymbol{t}^A \cdot \boldsymbol{v}^B \, dS = \int_{S_B} \boldsymbol{t}^B \cdot \boldsymbol{v}^A \, dS \tag{A.3}$$

Thereby the tractions are given by equation (11) applied to both excitation areas

$$\boldsymbol{t}^{A/B}(\tilde{r}, \omega) = -K^{A/B} \hat{H}(\omega) \, \text{grad}^2 \left( E^{A/B}(\tilde{r}) \right) \tag{A.4}$$

It should be mentioned that by distinguishing $K^A$ and $K^B$, the material properties relevant to wave excitation (that is the coupling constant (7)) may be different between $S_A$ and $S_B$. However, within each area they are assumed to be constant.

Corresponding transformations leading from (14) to (15) can be repeated. This results in

$$K^A \hat{H} \int_{S_A} [E^A \, \text{div}^2 \boldsymbol{v}^B] \, dS = K^B \hat{H} \int_{S_B} [E^B \, \text{div}^2 \boldsymbol{v}^A] \, dS. \tag{A.5}$$

We now introduce the spectra of both displacement fields again as (compare (16))

$$\boldsymbol{u}^{A/B}(\boldsymbol{r},\omega) := \widehat{H}(\omega)\boldsymbol{v}^{A/B}(\boldsymbol{r},\omega) \tag{A.6}$$

In Chapter 5.1 an averaged divergence $\boldsymbol{u}^B$ was defined, whereby the averaging was carried out over the excitation area of the complementary state A (see (18)). Now we do the same for both states B and A and indicate by additional indices which integration area and normalised energy density is involved

$$\overline{\text{div}\,u^B}^A := \int_{S_A} E_n{}^A \,\text{div}^2(\boldsymbol{u}^B)\,dS = \int_{S_A} E_n{}^A \,\text{div}^2\left(\widehat{H}\boldsymbol{v}^B\right)dS \tag{A.7}$$

$$\overline{\text{div}\,u^A}^B := \int_{S_B} E_n{}^B \,\text{div}^2(\boldsymbol{u}^A)\,dS = \int_{S_B} E_n{}^B \,\text{div}^2\left(\widehat{H}\boldsymbol{v}^A\right)dS \tag{A.8}$$

Putting all together we get a measurement reciprocity in the following form

$$\frac{\overline{\text{div}\,\mathbf{u}^B}^A}{\bar{E}_B K^B} = \frac{\overline{\text{div}\,\mathbf{u}^A}^B}{\bar{E}_A K^A} \tag{A.9}$$

The left-hand side of this equation refers to the state "B" which belongs to Fig. 16b and the right-hand side to the state "A" belonging to Fig. 16a. The Equation (A.9) refers to a measurement reciprocity for switching the excitation and detection positions when

- the excitation is done with a laser of given intensity distribution and
- the detection is done by a weighted area average of the surface displacement divergence, whereby the weight is the normalized energy density of the excitation in the reciprocal situation.

The measurement reciprocity is formulated in the frequency range but can be transformed back to time domain while keeping its form.

So far, we have aimed for a relationship that is as general as possible, allowing for the change in laser energy and energy distribution as well as different properties of the coupling material in both areas. Obviously, this can be specified to the more common situation of an identical laser spot and equal coupling factors by equating these values with each other:

$$\overline{\text{div}\,\mathbf{u}^B}^A = \overline{\text{div}\,\mathbf{u}^A}^B \tag{A.10}$$

A further simplification is possible if very small excitation points, i.e. point-shaped excitations, are considered. What is "point like" depends on the wavelength of the elastodynamic state. If the displacement divergence $\boldsymbol{u}^A$ does not changes significantly over the integration area $S_B$, this term can be put out of the integral

$$\overline{\text{div}\,u^A}^B := \int_{S_B} E_n{}^B \,\text{div}^2(\boldsymbol{u}^A)\,dS = \text{div}^2(\boldsymbol{u}^A)\int_{S_B} E_n{}^B\,dS = \text{div}^2(\boldsymbol{u}^A) \tag{A.11}$$

The divergence is now taken in the centre of the "point-like" excitation area $S_B$. As the same holds for $\boldsymbol{u}^B$ the measurement reciprocity reads now simply

$$\text{div}\,\mathbf{u}^B = \text{div}\,\mathbf{u}^A. \tag{A.12}$$

Finally, it must be mentioned that for the 2D problem of Chapter 5, all field variables do not depend on the position in the 3rd coordinate axis. Thus, the surface divergence simplifies to the surface strain that is the derivative of the in-plane displacement.